\documentclass[12pt]{iopart}

\usepackage{iopams}
\usepackage[dvips]{graphicx}

\begin{document}

\title{Nuclear equation of state and finite nucleon volumes}

\author{Jacek Ro\.zynek}
\address{National Centre for Nuclear Research, Ho\.za 69, 00-681 Warsaw, Poland}
\ead{rozynek@fuw.edu.pl}

\begin{abstract}
 It is shown how
the Equation of State (EoS) depends on nucleon properties inside Nuclear Matter (NM). We
propose to benefit from the concept of enthalpy in order to include volume corrections to
the nucleon rest energy, which are proportional to pressure and absent in a standard
Relativistic Mean Field (RMF) with point-like nucleons. As a result, the nucleon mass can
decrease inside NM, making the model nonlinear and the EoS softer. The course of the EoS
in our RMF model agrees with a semi-empirical estimate and is close to the results
obtained from extensive DBHF calculations with a Bonn A potential, which produce an EoS
stiff enough to describe neutron star properties (mass--radius constraint), especially
the masses of ``PSR J1614–2230'' and ``PSR J0348+0432'', known as the most massive ($\sim
2 M_\odot$)  neutron stars. The presented model has proper saturation properties,
including a good value of compressibility.
\end{abstract}
\maketitle

Taking into account the thermodynamic effects of pressure in finite volumes, we will
attempt to describe how the energy per nucleon $\varepsilon\!=\!M_A/A$ and the
pressure evolve with NM density $\varrho$ in a RMF approach
\cite{wa,ser,stocker,zm,GlenMosz}. The original Walecka version \cite{wa} of the
linear RMF introduces two potentials: a negative scalar $g_SU_S$ and a positive vector
$U_V\!=\!g_V(U_V^0,\mbox{\textbf{\emph{0}}})$ fitted to the nuclear energy
$\varepsilon$ at the equilibrium density $\varrho\!=\!\varrho_0$. The EoS for this
linear, scalar-vector ($\sigma,\omega$) RMF model \cite{wa,ser} matches a saturation
point with too large compressibility $K_A^{-1}\!=\!9\varrho^2\frac{d^2}{d\varrho^2}
\varepsilon\!\sim\!540MeV$ and is very stiff for higher densities where the repulsive
vector potential starts to predominate over the attractive scalar part. Nevertheless
RMF models produce, after the Foldy-Wouthuysen reduction, a good value of a large
spin-orbit strength \cite{wa} at the saturation density which depends on
$(g_VU^0_V\!\!-\!g_SU_S)$. The dynamics of potentials in the RMF approach are
discussed e.g. in \cite{ms2} in four specific mean-field models
\cite{wa,ser,stocker,zm}. In the ZM model \cite{zm} a fermion wave function is
re-scaled and interprets a new, density dependent nucleon mass. It starts to decrease
from $\varrho=0$ and at the saturation point $\varrho\!=\!\varrho_0$ reaches 85\% of
the nucleon mass $M_N$. But the nucleon mass replaced at the saturation point by a
smaller value would change the deeply inelastic Parton Distribution Function
\cite{Jaffe}, shifting the Bj\"{o}rken $x \propto (1/M_N)$. Such a shift means that
nucleons will carry 15\% less of the longitudinal momentum, which should be
compensated by an enhanced contribution from the meson cloud for small $x<0.3$ to
describe the EMC effect \cite{ms2,RW}. There is no evidence for such a huge
enhancement \cite{newdata} in the EMC effect for small x. Moreover, the nuclear
Drell-Yan experiments \cite{drell,ms2}, which measure the sea quark enhancement, have
been described \cite{jacek} with a small 1\% admixture of nuclear pions and the $M_N$
unchanged. Thus the deep inelastic phenomenology indicates that the change in the
nucleon mass at the saturation density is rather negligible. Another nonlinear
extension of the RMF model \cite{stocker,hansel} assumes self-interaction of the
$\sigma$-field with the help of two additional parameters fitted to
$K_A^{-1}\approx250MeV$ and an effective mass $M_N^*=M_N+g_SU_S$. These modifications
of the scalar potential give a softening of the EOS but are unstable for high
densities (\cite{ring}) and another nonlinear version has been proposed
(\cite{reinhard}). Modern RMF calculations \cite{Bielich,giai,bernardos,typel,meyer}
include the important Fock term of the nuclear mean field \cite{giai} and have
adjusted the EOS, fitting more meson fields ($\rho$ for an isospin dependence)
including the octet of baryons. The introduction of couplings of the meson fields to
derivatives of the nucleon field is a purely phenomenological approach \cite{typel}.

 There are existing works
\cite{exvol,Costa,Rocha} on excluded volume ``effects" in which nucleons with eigenvolume
$\Omega_N$ are treated as point-like particles in the available volume
$\Omega_A-A\Omega_N$; therefore vector and scalar densities are scaled respectively. The
properties of the ground state obtained\cite{exvol} are more realistic
 due to the additional repulsion from excluded volumes,  resulting in the reduction of the vector
 and scalar fields. In a similar approach Costa et al. discuss\cite{Costa} excluded volume
effects on nuclear matter properties and neutron star maximum masses and radii for the
different RMF models. They found that only nonlinear parameterizations retain the proper
NM compressibility. Using a different model Rocha et al.\cite{Rocha} introduce an
attractive phenomenological Nucleon-Nucleon (NN) potential in order to get saturation
properties for the repulsion generated by excluded nucleon volumes. However, a
non-relativistic approach to the NN interaction does not allow the use of this model for
dense nuclear matter inside neutron stars. There are also nuclear bag
models\cite{Kagiyama,Jennings,Hua} which discuss the density dependent bag constant and
radius in the nuclear equation of state. Kagiyama et al.\cite{Kagiyama} have obtained a
nucleon bag heavier and much smaller for higher densities; such a possibility will be
discussed here as rather unrealistic. The successful Quark-Meson-Coupling RMF
models\cite{Jennings,Hua} will be elaborated on and compared in section 1.2.

Our objective is to examine the main  approximations of the RMF, namely the constant
nucleon mass and the constant nucleon volume in the compressed nuclear medium.  The
nuclear EoS, in particular compressibility, depends on the NN interaction but also on
the compressibility $K_N^{-1}$ of quark matter confined inside nucleons. The novel
feature of our work is a direct volume contribution of pressure to the rest energy of
a nucleon, missed in previous works\cite{exvol,Costa,Rocha} on excluded volume effects
with the constant nucleon radius and also absent in Quark-Meson Coupling (QMC) models
\cite{Jennings,Hua}).

Any extended object inside a compressed medium (like a submerged submarine) needs
extra energy to preserve its volume. Thus from a ``deep" point of view, a finite
pressure correction should be taken into account in the RMF calculations. To describe
the possible dependence of the nucleon mass on the nucleon volume in a compressed
medium we adopt a nucleon bag model. For fixed pressure and  zero temperature it is
easy to show (see the first paragraph in the next section), that definitions of a
chemical potential $\mu$ or a Fermi energy, have the same energy balance as an average
single particle enthalpy. But enthalpy contains explicitly an interesting term, the
work of the nuclear pressure $p_H$ in the nuclear/nucleon volume, which will be
investigated. We consider NM for zero temperature, therefore our choice of enthalpy
corresponds to a Gibbs free energy with independent pressure $p_H$ in favor of a
Helmholtz free one (here an internal energy $M_N$) with the volume as an independent
variable. Our results are independent \cite{kumar} of this choice; like the expression
for the chemical potential $\mu$ in (\ref{enthsing}). The simplest ($\sigma,\omega$)
model \cite{wa,ser}, which is too stiff with point-like nucleons, will be used to
obtain clear conclusions. We will neglect nuclear pion contributions above the
saturation point. Dirac-Brueckner calculations show that the pion effective cross
section, in the reaction of two nucleons $N+N=N+N+\pi$, is strongly reduced at higher
nuclear densities above the threshold \cite{hm1} (also with RPA insertions to the self
energy of $N$ and $\Delta$ \cite{oset}). We restrict our degrees of freedom to
interacting nucleons. Further work for finite temperature is planned.
\section{Nuclear Enthalpy}
 \noindent At the beginning, let us consider effects generated by a volume of compressed
 NM starting with $A$ nucleons which occupy a volume $\Omega_A\!=\!A/\varrho$. They have to
perform the work $W_A\!=\!p_H\Omega_A$ necessary to keep the space $\Omega_A$ inside the
compressed NM against the nuclear pressure  $p_H$. The nuclear enthalpy at zero
temperature $T$ is given by:
\begin{eqnarray}
H_A\doteq M_A+W_A=M_A+A\ \!\frac{p_H}{\varrho}\label{enth} \\
p_H\doteq\!-(\partial M_A/\partial \Omega_{A})\mid_{T=0}
\end{eqnarray}
and contains, besides the nuclear mass as an internal energy, the necessary work.
Taking appropriate thermodynamical derivatives with respect to $A$, we get the following
relations between the chemical potential $\mu$ and the enthalpy for
$A\!\rightarrow\!\infty$:
\begin{eqnarray}
\mu\doteq\!(\partial M_A/\partial A)_{_{\Omega_A}}\!\!\equiv\!(\partial H_A/\partial
A)_{{p_H}}\!=\!\varepsilon\!+\frac{\!p_H}{\varrho}\!=\!H_A/A .\label{enthsing}
\end{eqnarray}
The same relation fulfils a nucleon Fermi energy with a Fermi momentum $\!P_F$
\begin{eqnarray}
E_F\!\doteq\!P_N^0(P_F\!)=\!(\partial M_A/\partial A)_{_{\Omega_A}}\!
 =\varepsilon\!+\!p_H/ \varrho=\!\mu\!; \label{HvH}
\end{eqnarray}
well-known as the  Hugenholtz-van Hove (HvH) relation\cite{kumar}, also proven in the
self-consistent RMF approach \cite{boguta}.
\subsection{Finite nucleon volumes in NM} \noindent There are two different
media inside nuclear matter. The hadron medium - where the NN interaction operates
inside $\Omega_{A-}=\Omega_A-A\Omega_N$ does not include nucleon volumes $\Omega_N$
filled with a confined quark matter with density $\varrho_N = N_q/\Omega_N$. Previous
work\cite{exvol,Costa} on the effect of excluding nucleon volumes assumed constant
nucleon volumes while the present work also discusses a decreasing volume $\Omega_N$
for a finite nucleon compressibility $K^{-1}_N\!>\!K^{-1}_A$. Let us justify the
possible modification of $K^{-1}_A$ for the saturation density. Pressure $p_H$ and
nuclear compressibility $K_A^{-1}$ are defined in the macroscopic volume
$\Omega_A=\Omega_{A-}+\Omega_N$ and thus also depend on nucleon properties such as
nucleon compressibility $K^{-1}_N$ which can be deduced from alpha scattering
experiments on protons. These estimates\cite{Mors} give for $K_N^{-1}=1.4\pm0.3$ GeV;
almost an order of magnitude greater then the nuclear compressibility $K_A^{-1}=215$
MeV measured\cite{Uchida} in alpha scattering on nuclear targets. The same
experiment\cite{Mors} found that possible effects of ``swelling" or ``shrinking" of a
nucleon in the nucleus must be very small. These two quantities at equilibrium can be
expressed as:
\begin{eqnarray}
\hspace{-20mm} K^{-1}_A|_{p_{H}=0}=9\frac{\partial p_H}{\partial
\varrho}=-9(A/\varrho^2)\frac{\partial p_H}{ \partial \Omega_{A}} \hspace{11mm}
K^{-1}_N|_{p_{H}=0}=9\frac{\partial p_H}{\partial \varrho_N}=-9(\Omega_{N}^2/N_q)
\frac{\partial p_H}{\partial \Omega_{N}} \label{Kdef}
\end{eqnarray}
Please note, that the ``external" pressure $p_H\doteq-\partial M_N/\partial
\Omega_{N}$ used in (\ref{Kdef})
 is identical with the nuclear pressure $p_H\doteq-\partial M_A/\partial \Omega_{A}$
appearing in (\ref{enth},\ref{enthsing}). The volume changes $\delta \Omega_{A},\delta
\Omega_{N}$ for the same increase of pressure are proportional to $K_{A},K_{N}$
respectively. For large $K^{-1}_N\!\gg \!K^{-1}_A$ we have (\ref{Kdef})
$\delta\Omega_{A}\!\gg\!(\Omega_N/\varrho)^2 A\delta\Omega_{N}>A\delta\Omega_{N}$ and
actually the pressure will mostly squeeze the space between the nucleons, $\Omega_{A-}
= \Omega_A - A\Omega_N$. In the limit of constant $\Omega_N$, $p_H$ can be written as:
\begin{eqnarray}
p_H\!\!\mid_{_{\Omega_N}}\doteq -\partial M_A/\partial
\Omega_{A}\!\mid_{_{\Omega_N}}=-\partial M_A/\partial \Omega_{A-}\!\mid_{_{\Omega_N}}.
\label{press}
\end{eqnarray}
The total enthalpy $H^T_A\!\doteq \!H_{A-}+A(H_N\!-\!M_N\!)$ and using
(\ref{enth},\ref{enthsing},\ref{press},\ref{enthnuc}) we arrive at the HvH relation
(\ref{HvH}) for NM containing  fixed sized nucleons:
\begin{eqnarray}
H^T_A\!/A\!\mid_{_{\Omega_N}}\!=\!\varepsilon\!-\!{(\partial M_A/
\partial \Omega_{A})\!}_{A,\Omega_N}(\Omega_{A-}/A+\Omega_N)=\!\varepsilon\!+\!p_H\!/\!\varrho=\!E_F .\label{enthtot}
\end{eqnarray}
In the next section arguments for a constant nucleon size in compressed NM will be
discussed.
\subsection{The nucleon mass in the Bag model in NM}
In order to consider nucleon volume corrections in the RMF we have to introduce a model
for an extended nucleon. Let us take the simplest possibility, the Bag Model. In this
model we will discuss whether the nucleon mass $M_N$ as the internal energy or rather a
total nucleon energy $H_N$ should be, eventually, constant - independent of the density
inside the compressed medium. Such a question is absent in the standard RMF where
nucleons are point-like with a constant mass $M_N$ independent of the pressure inside NM.
But nucleons themselves are extended. However, in a compressed nucleon, partons (quarks
and gluons) have to do work $W_N=p_H\Omega_N$ to keep the space $\Omega_N$ for a nucleon
``bag". This will involve functional corrections to the nucleon rest energy, dependent on
the external pressure with a physical parameter - the nucleon radius $R$. Others
modifications connected with the finite volume of nucleons, such as correlations of their
volumes, will be neglected. We introduce a nucleon enthalpy $H_N$ with the nucleon mass
$M_{pr}$ possibly modified in the compressed medium
\begin{eqnarray}
H_N(\varrho)\doteq M_{pr}(\varrho)+p_H\Omega_N \ \ with \ \ H_N(\varrho_0)=M_N, \ \
\label{enthnuc}
\end{eqnarray}
 as a ``useful" expression for the total rest energy of a nucleon ``bag".

\noindent Describing nucleons as bags, pressure will influence their surfaces
\cite{Koch,Hua,bag,Kap,Jennings}. Finite pressure corrections to the mass can not be
described clearly by perturbative QCD \cite{brown}. Let us discuss the relation
(\ref{enthnuc}) in the simple bag model where the nucleon in the lowest state of three
quarks is a sphere of volume $\Omega_{N}$. In vacuum its energy $E^0_{Bag}$ \cite{MIT} is
a function of the radius $R_0$ with phenomenological constants - $\omega_0$, $Z_0$
\cite{Hua} and a bag ``constant" $B(\varrho)$:
\begin{eqnarray}
E^0_{Bag}\!(R_0)\!\!&=&\frac{3\omega_0-Z_0}{R_0}+\frac{4\pi}{3}B(\varrho\!=\!0)R_0^3\propto~\!1/R_0.
\label{bag}
\end{eqnarray}
The condition
\begin{eqnarray}
p_B=-\left(\partial E^0_{Bag}/\partial \Omega_{N}\right)_{surface} =0
\label{pressured2}
\end{eqnarray}
for the pressure inside a bag in equilibrium, measured on the surface, gives the relation
between $R_0$ and $B$, used at the end of (\ref{bag}). $E_{Bag}^{0}$ fits to the nucleon
mass at equilibrium density $\varrho_0$ where $p_H\!=\!p_B\!=\!0$.
 ($E^0_{Bag}$ differs from $M_N$ by the c.m. correction \cite{Jennings}).

In a compressed medium, the pressure generated by free quarks inside the bag \cite{MIT}
is balanced at the bag surface not only by the intrinsic confining ``pressure"
$B(\varrho)$ but also by the nuclear pressure $p_H$; generated e.g. by elastic collisions
with other hadron \cite{Koch,Kap} bags, also derived in the QMC model in a medium
\cite{Hua}. In a medium, internal parton pressure $p_B$ (\ref{pressured2}) inside the bag
is equal (cf. \cite{Hua}) on the bag surface to the nuclear pressure
\begin{eqnarray}
p_H\!=p_B\!\!&=&\!\!\frac{3\omega_0-Z_0}{4\pi
R^4}-\!B(\varrho)~~\rightarrow~~4\pi(B(\varrho)\!+\!p_H\!)R^4\!=\!3\omega_0-Z_0
\label{pressure}
\end{eqnarray}
and we get the radius $R$, depending from a sum $(B\!+\!p_H)$:
\begin{eqnarray}
R(\varrho)\!&=&\!\left[\frac{3\omega_0-Z_0}{4\pi (B(\varrho)+p_H(\varrho))}\right]^{1/4}.
 \label{rsolution}
\end{eqnarray}
Thus, the pressure $p_H(\varrho)$ between the hadrons acts on the bag surface similarly
to the bag ``constant" $B(\varrho)$. A mass $M_{pr}$
 can be obtained from (\ref{bag},\ref{pressure}):
\begin{eqnarray} \hspace{-20mm}
M_{pr}(\varrho)\!\!=\frac{3\omega_0-Z_0}{R}+\frac{4\pi}{3}B(\varrho)R^3\!=\!\frac{4}{3}\pi\!
R^3\!\left[4(B+p_H)\!-\!{p_H}\right]\!= \!E_{Bag}^{0}\frac{R_0}{R}\!-\!p_H\Omega_{N}.
\ \  \label{massbag}
\end{eqnarray}
 The scaling factor $R_0/R$ comes from
the well-known model dependence (\ref{bag}) ($E_{bag}^0\!\propto\!1/R_0$) in the
spherical bag \cite{MIT}. This simple radial dependence is now lost in (\ref{massbag})
and the pressure dependent correction to the mass of a nucleon given
 by the product $p_H \Omega_N$, identical with the work $W_N$ in
(\ref{enthnuc}), is responsible for this and disappears in the following expression
for the nucleon enthalpy
\begin{eqnarray}
H_N(\varrho)=M_{pr}\!+\!p_H\Omega_{N}=({R_0}/R)E_{Bag}^{0}\propto1/R(\varrho) .
\label{enthbag}
\end{eqnarray}

The nucleon radius $R(\varrho)$ reflects the scale of confinement of partons. Generally,
for increasing $R(\varrho)$, $H_N(\varrho)$ (\ref{enthbag}) (and $M_{pr}$) decreases,
thus part of the nucleon rest energy is transferred from a confined region $\Omega_{N}$
to the remaining space $\Omega_{A-}$ (\ref{enthtot}). This scenario is rather impossible
at the equilibrium density where nucleon compressibility is positive\cite{Mors}. For
decreasing $R$, $H_N$ increases and the mass $M_{pr}\!=\!H_N\!-p_H\Omega_N$ changes
(\ref{enthnuc}) according to the EoS. When NN interactions do not change the rest energy
$H_N(\varrho)\!=\!M_N$ of partons confined inside nucleons, the radius $R(\varrho)$ is
constant (\ref{enthbag}). It requires work $W_N$ to keep a constant volume $\Omega_{N}$
at the expense of the nucleon mass $M_{pr}$ (\ref{massbag}) and is obtained
(\ref{rsolution}) for the constant effective pressure
$B_{eff}\!\!=\!B(\varrho)\!+\!p_H(\varrho)\!=\!B(\varrho_0)$. The
$B(\varrho)\!=\!B(\varrho_0)\!-\!p_H$ gradually decreases with pressure and disappears
 for $p_H(\varrho)\!\!=\!\!B(\varrho_0)\!\simeq60$ MeVfm$^{-3}$ \cite{MIT}; in favor of strongly
correlated colored quarks in the de-confinement phase when
$\varrho\!\approx\!(0.5\!-\!0.6)$ fm$^{-3}$. For constant  $M_N$ the bag constant
decreases more slowly and disappears for $p_H\simeq100$ MeVfm$^{-3}$.

The internal pressure $B(\varrho)$, just as $p_H(\varrho)$ (generated by meson
exchanges), originates \cite{Bub} from interactions of quarks. Therefore, by increasing
$p_H(\varrho)$ we can expect a corresponding response in $B(\varrho)$. Actually, when
pressure $p_H$ in NM is not taken into account ($p_H=0$ in (\ref{rsolution})) the nucleon
radius $R$, in the QMC model \cite{Jennings}, increases in NM. However, in the updated
\cite{Hua} QMC model, which takes into account $p_H$ contributions to the bag radius, a
dependence of $R(\varrho)$ on density is a specific property of a particular nuclear
model or EoS. In particular, for the ZM model \cite{zm}, which has the realistic value of
$K_A^{-1}\!\simeq\!225$ MeV, the nucleon radius remains almost constant up to density
$\varrho\!=\!10\varrho_0$ (the volume corrections (\ref{massbag}) to the nucleon mass are
absent here). However, for the too stiff EOS of the linear $(\sigma,\omega)$ model, they
\cite{Hua} observe a strong increase of the nucleon radius up to density
$\varrho=2\varrho_0$. Such an increase of the radius would diminish the total rest energy
$H_N$ (\ref{enthbag}) and the nucleon mass (\ref{massbag}), making the EOS substantially
softer - as a consistent feedback. Besides, it has been shown in a Global Color Symmetry
Model (GCM) \cite{bag}, that a decrease of $B(\varrho)$ from the saturation density
$\varrho_0$ up to $3\varrho_0$ by $~60$ MeVfm$^{-3}$, is accompanied by a similar
increase of the pressure $p_H$.
Summarizing, the sum $B(\varrho)\!+\!p_H(\varrho)$ weakly depends on density in GCM or
QMC models with a reasonably stiff EOS, thus the bag radius remains approximately
constant (\ref{rsolution}). This justifies the choice of constant total rest energy
$H_N$, unchanged by increasing NN repulsion, just opposite to the case with constant
$M_N$ which requires an increase of the rest energy $H_N(\varrho)$
(\ref{enthnuc},\ref{enthbag}) with decreasing $R(\varrho)$. \vspace*{-0mm}

\section{Results and Discussion}
\subsection{The nuclear compressibility - a start of the EOS}
Before presenting the results for the EOS, let us consider the nuclear compressibility
$K_A^{-1}$ at the starting point at $\varrho=\varrho_0$. A recent self-consistent
calculation presented in the thorough review \cite{Jirina} of the nuclear compressibility
shows that the density dependence of the surface diffuseness of a vibrating nucleus plays
a role in determining the nuclear compressibility. Therefore, we can expect that density
dependent properties of nucleons embedded in NM will also modify $K_A^{-1}$. There are
two cases when we can easily derive the nucleon compressibility from
(\ref{Kdef},\ref{massbag}): at constant mass $M_N$ - case \textbf{A} (and density
$\varrho_0$) or at constant volume $\Omega_N$ - case \textbf{B}:
\begin{eqnarray}
\hspace{-25mm} K^{-1}_N\!\!\mid_{_{M_N,R\rightarrow R_0}}\!=\!-3\Omega_N^2\frac{\partial
[M_N(R_0/R-1)/\Omega_N]}{\partial \Omega_N}\!=\!M_N\!\simeq\!940\!\!\texttt{ MeV}
\hspace{7mm}\texttt{or} \hspace{7mm}K^{-1}_N\!\!\mid_{_{\Omega_N}}\!\rightarrow\!\infty.
\label{estim}
\end{eqnarray}
We have already mentioned that the experimental estimate\cite{Mors}, using the quark sum
rule, gives $K^{-1}_N\!=\!(1.4\pm0.3)$ GeV, in between our two estimates in
Eq.(\ref{estim}); while other theoretical models (refs. in \cite{Mors}) give a broad
range of $K^{-1}_N\!=\!(0.6-3.0)$ GeV estimates. There is an open question whether the
nucleon changes its size and/or mass in medium. Therefore,  cases A and B will be
discussed and compared.

In case \textbf{A} the nucleon mass is constant; according to Eq.\ref{massbag}
 the nucleon radius will decrease in pressure (density) in order to provide the volume energy;
 see Fig.1 - right panel. However the
volume changes depend on the relative compressibility $K^{-1}_{rel}=K_N^{-1}/K_A^{-1}$.
The upper limit of a corrected nuclear compressibility $K_{A_\Omega}^{-1}$ - which takes
into account excluded volume effects, can be obtained in the limit of
$K_N^{-1}/K_A^{-1}\gg1$ when the nuclear compressibility will be scaled like pressure in
(\ref{press}):
\begin{eqnarray}
K_{A_\Omega}^{-1}\mid_{p_H=0}\simeq K_{A}^{-1}\mid_{p_H=0}/(1-\varrho\Omega_N)\approx
1.15K_{A}^{-1}\mid_{p_H=0}
\end{eqnarray}
However, this is only an upper limit, thus a realistic value of the relative
compressibility $K^{-1}_{rel}$ is needed here. Anyway, a small increase of nuclear
compressibility is expected in the case of the constant nucleon mass.

In case \textbf{B}, the nucleon radius is fixed (the arguments were given at the end of
section $1.2$), thus a volume correction will diminish (\ref{massbag})
 the effective nucleon mass similarly to the change made by the scalar part $g_SU_S$ of
\begin{eqnarray}
M^*_{pr}(\varrho)\!&=&\!M_N +g_S U_S-p_H(\varrho)\Omega_N, \ \ \ \ \ \varrho\geq\varrho_0
\label{efmass}
\end{eqnarray}
the mean field potential; see Fig.1 - left panel. This will involve an additional
nonlinear term $p_H\Omega_{N}$ in the expression for the energy $\varepsilon$ in the
RMF \footnote{In the realistic Hartree-Fock \cite{giai} version of RMF, the volume
correction $p_H(\varrho)\Omega_N$ will also modify the following ratio:
\begin{eqnarray}
\frac{M_{pr}^*}{\sqrt{\mbox{\emph{\textbf{P}}}_N^2\!+\!{M_{pr}^{*2}}}}\!=\!\frac{M^*_N-p_H(\varrho)\Omega_N
}{\sqrt{\mbox{\emph{\textbf{P}}}_N^2\!+\!{(M_N^*-p_H(\varrho)\Omega_N)^2}}}\simeq\frac{M_N^*}
{\sqrt{\mbox{\emph{\textbf{P}}}_N^2\!+\!{M_N^{*2}}}}\ .  \label{ratio}
\end{eqnarray}
But not far from the saturation density $\rho_0$ the volume correction can be neglected
in this ratio.}.
 The difference between compressibility $K_{A_\Omega}^{-1}$ - with, and $K_A^{-1}$
-without volume correction to energy $\varepsilon$ can be written in case \textbf{B}
as:
\begin{eqnarray}
\hspace{-25mm}
K_A^{-1}\!-\!K_{A_\Omega}^{-1}\!=\!9\varrho^2\frac{\partial^{2}(p_H\Omega_{N})}{\partial\varrho^2}\!=\!
9\varrho^2\frac{\partial ^2}{\partial \varrho^2}\!\left[\frac{r\varrho}{1\!-
\!r}\frac{\partial \varepsilon}{\partial \varrho}\right]\!=\!
\frac{9r\varrho^2}{1\!-\!r}\!\left[f(\varrho)\frac{d\varepsilon}{d\varrho}+\!
\left(\frac{6\!-\!5r}{1\!-\!r}\!\right)\!\frac{
d^2\varepsilon}{d\varrho^2}\!+...\right] \label{Keq}
\end{eqnarray}
where the parameter $r=\varrho\Omega_N$ and $f(\varrho)$ is an unknown regular function
at the equilibrium density which multiplies the first derivative $\varepsilon'(\varrho)$
vanishing at $\varrho_0$. The last term in (\ref{Keq}) with the third derivative is
omitted. Consequently, keeping only the term with the second derivative
$\varepsilon''(\varrho)$ which is proportional to $K_{A_\Omega}^{-1}$, we finally obtain:
\begin{eqnarray}
K_{A_\Omega}^{-1}\mid_{p_H=0}&\cong&\left[\frac{(1-\varrho
\Omega_N)^2}{1+4\varrho\Omega_N(1-\varrho\Omega_N)}\right] K_A^{-1}\simeq\frac{1}{2}
K_A^{-1}\mid_{p_H=0} \label{K}
\end{eqnarray}
that the compressibility $K_{A_\Omega}^{-1}$ which includes the volume corrections is
smaller by the density dependent factor shown above in (\ref{K}). The typical
saturation density with $\Omega_N\approx1$ provides a small
$\varrho\Omega_N\approx1/6$ for $R_0=0.6$fm. Consequently $K_{A_\Omega}^{-1}\approx
K_A^{-1}/2$ (inside uniform NM). It is quite a remarkable reduction which proves how
important volume corrections in the vicinity of equilibrium are in the case \textbf{B}
(constant $\Omega_N$), although energy corrections disappear with pressure for
$\varrho_0$. Note that the NM compressibility obtained in the the basic
Hartree\cite{wa} and Hartree Fock \cite{giai} calculations give
$K_A^{-1}\approx(460-560)$ MeV; when it is reduced by a factor 1/2, it arrives at a
proper value.
\begin{figure}
\hspace{.4cm} \includegraphics[height=6.cm,width=8.3cm]{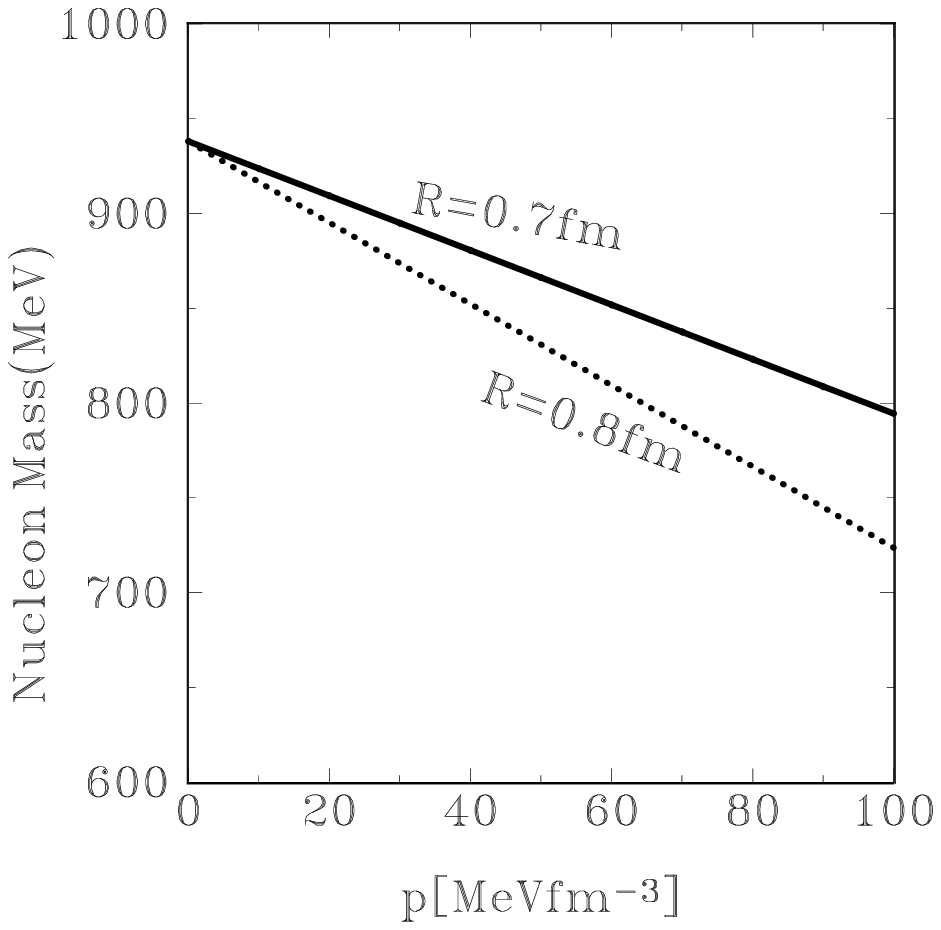} \caption{right
panel: Case \textbf{A} - pressure dependent nucleon radius $R$ for two different initial
values of $R_0$ at the equilibrium density. Left panel: Case \textbf{B} - the nucleon
mass $M_{pr}$ as a function of NM pressure for two nucleon radii R=0.7fm,0.8fm.}
\label{fig:enernew}
\end{figure}
\begin{figure}
\vspace{-9.4cm} \hspace{8.3cm}
\includegraphics[height=7.3cm,width=7.8cm]{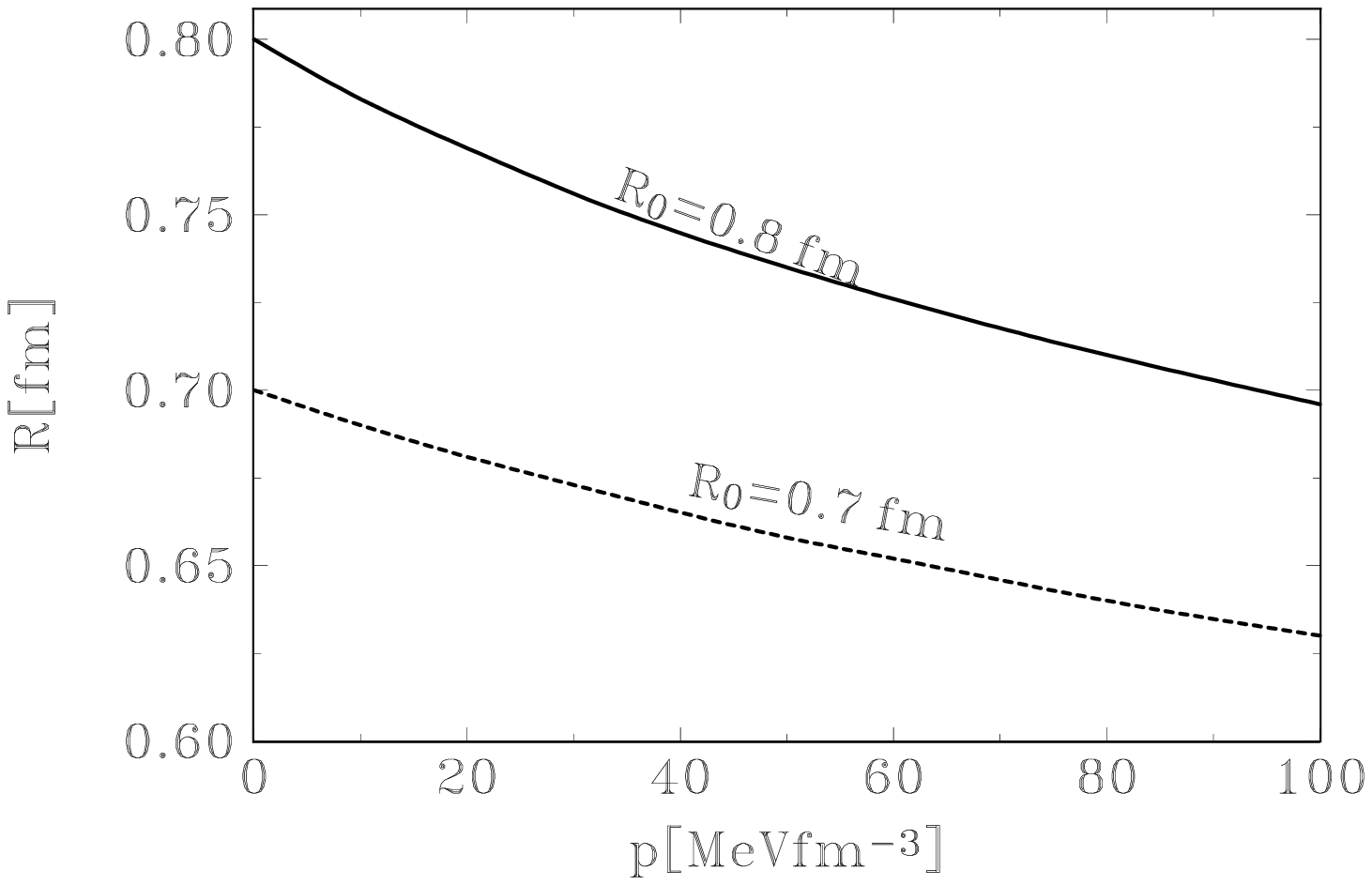}
\label{fig:radius} \vspace{.6cm}
\end{figure}

\subsection{The EoS} The results for energy and pressure of NM in case \textbf{A} remain practically
unchanged in comparison with the standard classical ($\sigma,\omega$) RMF calculation
with a stiff EOS and large compressibility $K_A^{-1}\simeq600MeV$.

At the end of section $1.2$ we presented arguments for the second scenario \textbf{B} -
constant nucleon volume $\Omega_N$ (hence constant rest energy $H_N(\varrho)=M_N$),
regardless of pressure. In that case we apply the following formulas
(\ref{enthnuc},\ref{press}) for the nucleon mass $M_{pr}$ modified inside NM with
pressure, which is now given by:
\begin{eqnarray}
p_H\!=\!-{(\partial M_A/
\partial \Omega_{A
})}_{A,\Omega_{N}}\!&=&\!\varrho^2 {{\varepsilon}^{'}}(\varrho)\!/(1\!-\!\varrho
\Omega_N\!). \label{enth1}
\end{eqnarray}
To carry out calculations we combine the $M_{pr}$ dependence (\ref{massbag},\ref{efmass})
of the pressure $p_H$ at the constant nucleon radius $R\!\!=\!\!R_0$ with the following
linear RMF equations \cite{wa,ser} for the energy $\varepsilon$ in terms of the effective
mass $M_{pr}^*$:
\begin{eqnarray}
\varepsilon\!=\!&g&\!\!\!\!\!_V \frac{U^0_V}{2}
\!+\!\frac{C_2^2}{\varrho}(M_{pr}\!-\!M_{pr}^*)^2\!\!+
\!\frac{\gamma}{\varrho}\!\!\int_0^{P_F}\!\frac{d^3\!\mbox{\emph{\textbf{P}}}\!_N}{(2\pi)^3}\sqrt{\mbox{\emph{\textbf{P}}}_N^2\!+\!{M_{pr}^{*2}}}
\nonumber\\
M^*_{pr}\!\!&=&\!M_{pr}-\frac{\gamma}{2C_2^2}\int_0^{P_F}\!\frac{d^3\!\mbox{\emph{\textbf{P}}}\!_N}{(2\pi)^3}
\frac{M^*_{pr}}{\sqrt{\mbox{\emph{\textbf{P}}}_N^2\!+\!M^{*2}_{pr}}}. \label{eq}
\end{eqnarray}
$\gamma$ denotes the level degeneracy and there are two (coupling) constants: a vector
$C_v^2$ and a scalar $C_s^2$, which were fitted at two different saturation points in NM:
$S_1$\cite{ser} with $\varrho_0\approx 0.149$ fm$^{-3}$ and $S_2$\cite{wa} with
$\varrho_0\approx 0.193$ fm$^{-3}$  -- see the figure caption. In formula (\ref{eq})
$2C_1^2\!=\!C_v^2/M_N^2$, $2C_2^2\!=\!M_N^2/C_s^2$ with $g_VU^0_V\!=\!2C_1^2\varrho$,
$g_SU_S\!=\!M_{pr}\!-\!M^*_{pr}$. Now the finite pressure corrections to $M_{pr}$
(\ref{enth1}) convert the recursive equations (\ref{eq}) above the $\varrho_0$ to a
differential-recursive set of equations, taking the general form:
\begin{equation}
f(\varepsilon (\varrho),\varepsilon^{'}(\varrho))=0 ~for ~\varrho\geq\varrho_0.
\label{eq2}
\end{equation}
Note that (\ref{eq}) is obtained from the energy--momentum tensor for the model
Hamiltonian with a constant nucleon mass \cite{wa}. Here, we assume that the same
equation with mass $M_{pr}$ is satisfied in compressed NM. It should be a good
approximation, at least not very far from the saturation density. The corresponding
HvH relation (\ref{enthtot}) for the total nuclear enthalpy:
$H_A^T/A=E_F\!=\!\varepsilon\!+\!p_H/\varrho$, connects the Fermi energy with the
nuclear pressure $p_H$ and is met with $(0.1\!-3.)\%$ numerical accuracy; worse for a
higher density, ensuring the convergence of our numerical procedure (\ref{eq2}).
\begin{figure}[t]
 \vspace{-4mm} \hspace*{15mm}
\includegraphics[height=11.cm,width=14.4cm]{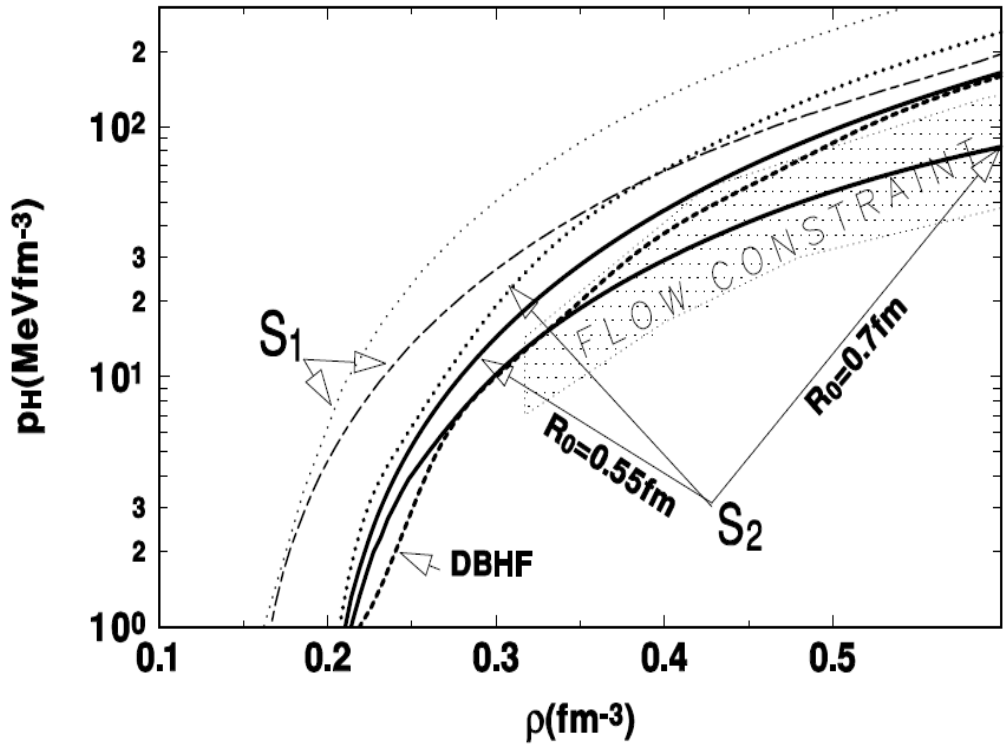}
\caption{\vspace{-1mm}Dotted lines show the pressure for a constant nucleon mass $M_N$
(case \textbf{A}) as a function of the density for two different parameterizations of the
($\sigma,\omega$) RMF model: $S_1$\cite{ser} with $P_F=1.30$ fm$^{-1}$
 and $S_2$\cite{wa} with $P_F=1.42$
fm$^{-1}$. The long dashed line shows  results for case \textbf{B} for constant nucleon
radius  $R_0=0.7$ fm and $S_1$ parametrization. Results for set $S_2$ with $R_0=0.55$ fm
or $R_0=0.7$ fm, marked as solid lines, are indicated by arrows. The phase transition to
quarks, for $p_H \gtrsim \!B(\varrho_0)\approx60$ MeV, is not considered (large $B$
limit).
 The area indicated by
``flow constraint'' taken from \protect\cite{pawel} determines the allowed course of the
EoS, using an analysis which extracts from the matter flow in heavy ion collisions from
the high pressure obtained there. The DBHF \cite{bonn} calculations with the Bonn $A$
interaction are shown as short dashes.} \label{main} \end{figure}

Linear ($\sigma\!-\!\omega$) models \cite{wa,ser} with constant mass $M_N$ produce a
too stiff EoS, see Fig.\ref{main}. Our results, which take into account nucleon
volumes, are compared with a semi-empirical flow constraint \cite{pawel} from heavy
ion collisions and indeed they correct the EOS making it much softer. Fig.\ref{main}
shows a good course of the EoS in NM for \{$R_0=0.7$ fm, set $S_2$\} up to a density
$\varrho=0.6$ fm$^{-3}$. In fact, around this density, (partial) de-confinement is
expected, which will change the EoS above a phase transition \cite{quarkmatter}. For
\{$R_0=0.55$ fm, set $S_2$\} the EoS is relatively stiffer, slightly above the semi
experimental flow constraint. However, it is a good candidate to investigate closely
compact stars \cite{last} in the case when hyperons ``soften" \cite{hansel,strange}
the EoS further. In Fig.\ref{main}, it can be seen that both results for set $S_2$ are
rather close to the DBHF results \cite{bonn} which produce an EoS able to describe
\cite{NSTARS} the mass of ``PSR J1614–2230'' or ``PSR J0348+0432'' stars\cite{pulsar}
(for $R_0=0.7$ fm slightly below the DBHF for higher densities).  It is worth
mentioning that in the DBHF model there is additional correction \cite{bonn} from the
self-energy, which also diminish the nucleon mass with density. Alternatively, for an
additional softening of the EOS
 the $S_1$ parametrization with our corrections (dashed
line) can be considered.

In our model a volume part $p_H\Omega_N$ (\ref{enth1}) originating from the constant
total rest energy $H_N\!=\!M_N$ (\ref{enthnuc},\ref{enthbag}), effectively diminishes
the nuclear compressibility, changing its value from the unrealistic
$K_A^{-1}\!=\!540$ MeV \cite{wa} (set $S_2$) to the reasonable $K_A^{-1}\!=\!328;\
\!252;\ \!172$ MeV obtained in our model for $R_0\!=\!0.6;\ \!0.7;\ \!0.8$ fm
respectively, which is in good agreement with our estimate (\ref{K}). Therefore, the
nucleon volume $\Omega_N$ is an important physical factor which strongly reduces the
compressibility (\ref{K}) and stiffness in case \textbf{B}.
\section{Conclusions}
\vspace{-0mm} \noindent It has been shown how the nucleon properties may affect the
nuclear compressibility and EoS in the RMF model. Similar changes are expected in the
Hartree-Fock version \cite{giai,ring} because the nonlinear corrections to energy are
almost the same (\ref{ratio}).

In case \textbf{A}, with a constant nucleon mass and a decreasing radius (\ref{massbag}),
the changes in the EoS are negligible. However, because the nucleon radius decreases, the
de-confinement transition is shifted from $p_H\approx60MeVfm^{-3}$ (case \textbf{B}) to
the higher density $\varrho>0.6$fm$^{-3}$ - where the bag constant finally disappears
(\ref{pressure}) at a pressure $p_H\approx100MeVfm^{-3}$.

In case \textbf{B} the nucleon radius is constant and the nucleon mass (\ref{massbag})
$M_{pr}(\varrho)$ occurred to be a pressure functional. Such a weak dependence of $R$ on
$\varrho$ is consistent\cite{Hua} with the phenomenological EOS. In this case we have
found that the total rest energy $H_N$ of the nucleon is independent of density
(\ref{enthbag}), therefore the nucleon mass (\ref{massbag}) decreases with $\varrho$ and
pressure. It effectively corresponds to nonlinear modifications of a scalar potential.
Not accidentally, in the widely used standard \cite{Bielich,hansel,ring,reinhard} RMF
model with point-like nucleons the good compressibility is fit by nonlinear changes of a
scalar mean field, using two additional parameters (also introducing the density
dependent meson coupling constants \cite{typel}).  Our results indicate that the so
called ``excluded volume corrections" are the real source of nonlinear modifications in
the nuclear medium with finite pressure, which enables calculations with a smaller number
of free parameters. The compressibility \cite{pie} is lowered to an acceptable value,
giving a
 good course of the EoS for higher densities.

So far, the experimental compressibility $K_N^{-1}$ indicates that we are in between
 these two scenarios. The
presented model is suitable for studying heavy ion collisions and neutron star
properties (mass--radius constraint); especially the most massive known neutron
stars\cite{pulsar} recently discovered and we plan to include the Fock term and the
octet of baryons in future calculations.
\section*{Acknowledgment} \vspace{-2mm}
I would like to express my gratitude to J.R. Stone for our fruitful discussions and her
helpful remarks. This work is supported by the National Science Center of Poland, grant
DEC-2013/09/B/ST2/02897.

\end{document}